\documentclass[runningheads]{llncs}
\usepackage[T1]{fontenc}
\usepackage{graphicx}
\usepackage{hyperref}
\usepackage{float}
\usepackage[section]{placeins} 

\usepackage{tikz}
\usetikzlibrary{arrows.meta,positioning,shapes,fit}
\usepackage{amsmath,amssymb,amsfonts}
\usepackage{algorithm2e}

\def\BibTeX{{\rm B\kern-.05em{\sc i\kern-.025em b}\kern-.08em
    T\kern-.1667em\lower.7ex\hbox{E}\kern-.125emX}}

\usepackage{url}
\usepackage{hyperref}

\begin{document}
\title{Scaling Homomorphic Applications in Deployment}

\author{Ryan Marinelli\inst{1}\orcidID{0009-0001-7279-3156}
\and Angelica Chowdhury\inst{2}\orcidID{0009-0008-1985-1047}}
\authorrunning{Marinelli and Chowdhury}
\institute{%
\inst{1}University of Oslo, Department of Informatics, Gaustadalléen 23B, 0373 Oslo\\
\email{ryanma@ifi.uio.no}\\
\url{https://www.mn.uio.no/ifi/english/research/groups/sec/index.htm}
\and
\inst{2}Independent Researcher\\
\email{angelica.chowdhury2017@gmail.com}
}


%
\maketitle               
\begin{abstract}
In this endeavor, a proof-of-concept homomorphic application is developed to determine the production readiness of encryption ecosystems. A movie recommendation app is implemented for this purpose and productionized through containerization and orchestration. By tuning deployment configurations, the computational limitations of Fully Homomorphic Encryption (FHE) are mitigated through additional infrastructure optimizations.

\end{abstract}

\section{Introduction}
\textbf{Homomorphic Encryption} is a type of encryption that allows for computations on encrypted data. This capability is significant as it provides an additional layer of protection for end users. For instance, man-in-the-middle attacks are among the most pervasive cyber threats, allowing an interloper to intercept messages between the client and the server\cite{owasp_man_in_the_middle_attack}. The fundamental purpose of encryption is to ensure that only the sender and the intended receiver can access the message. If the data is encrypted throughout the system, even during computation as in the case of homomorphic encryption, the threat landscape shrinks in comparison to using other conventional encryption schemes. The main paradigms of encryption have focused on encryption at rest and encryption in transit. These occur at different layers  of the OSI model. The most significant being the Network Layer and Presentation Layer for encryption \cite{russell2013osi}.

When it comes to information at rest, storing information at rest is becoming increasingly important, especially considering cloud infrastructure. Ristenpart et al. \cite{ristenpart2009hey} found that the way providers distribute resources can be problematic. Providers share hardware across their customer base, which allows them to benefit from compute flexibility; however, these shared resources have been shown to leak information between environments creating an attack vector. 

Regarding data in transit, this area remains a prominent point of concern. HTTPS and related protocols depend on certificates, which have sometimes proven troublesome from a security perspective. An infamous incident involving DigitNotor saw the issuance of hundreds of fake certificates, highlighting drastic variations in certificate credibility \cite{hkcert_diginotar_2011}. For example, Mozilla and Chrome no longer recognize certificates issued by Entrust\cite{mozillaDevSecPolicy}.

Homomorphic encryption offers mitigation for these concerns. In cloud environments, it provides protection against data leakage and enhances the integrity and confidentiality of services offered by cloud providers and multi-cloud architectures, which are increasingly utilized by developers. For data in transit, it offers similar benefits as other forms of encryption. Additionally, if an application processes data in an encrypted state, even in the event of leaks or vulnerabilities, the confidentiality of the data remains intact, providing a unique advantage. Essentially, if one were using a cloud service to power an application, they would only need to send encrypted data and leverage the compute resources. Even if there were a leak, the confidentiality of the data would be maintained. In the current state, the data would only be encrypted in transit and decrypted for processing to use the compute of the potentially vulnerable environment.

 \section{Literature Review}
Jian et al. conducted an experiment in applying deep reinforcement learning to enhance Kubernetes clusters \cite{https://doi.org/10.1002/spe.3284}. The primary focus of their work is load balancing. They define the state space as the resource utilization of each node in a cluster, computed as a composite metric of CPU, memory, network, and disk usage. The action space is defined as selecting a node on which a pod executes. The reward function is designed to reduce imbalance and increase average resource utilization. One strength of this work is its exploration of the dynamics across multiple clusters and various load balancing strategies, incorporating a diverse set of metrics into the reward function.

The work could be extended by investigating the configuration of each cluster and the scaling of pods within a cluster, rather than load balancing across different clusters. This avenue of investigation has helped to motivate the present endeavor by \textbf{determining that the unit of analysis should be focused on intra-cluster rather than inter-cluster dynamics.}

Regarding the state of homomorphic encryption and AI, Hamza provides comprehensive coverage of the current landscape \cite{10299436}. In his work, Hamza compares different algorithms and libraries. One of the libraries discussed is “concrete.” He notes that it has excellent documentation and an active community. This is exemplified by projects built on “concrete,” specifically “concrete-ml” \cite{ConcreteML}. However, it is argued that “concrete” is not scalable despite being well supported. To mitigate this concern, the present endeavor will focus on applying “concrete-ml” in a production scenario and investigating infrastructure optimizations to address these criticisms. By addressing scalability concerns, \textbf{this work aims to support the most user-friendly libraries for FHE and help popularize these tools as the ecosystem matures.}
 
\section{Methodology}
In this research endeavor, an application is developed to demonstrate the feasibility of a production-ready implementation of fully homomorphic encryption and to further strengthen deployment strategies. In this case, a movie recommendation app is developed using FHE based inference as a proof-of-concept. The application is a Flask application\cite{flaskDocs} with a productionized server\cite{waitressDocs}. The MovieLens dataset is utilized\cite{harper2015movielens} with the concrete-ml library, which provides a scikit-learn interface \cite{ConcreteML}. A naive logistic regression model is used to suggest a movie from a pool of 50 films in the dataset. The inference is run on a trained model that uses FHE. The entire pipeline of using FHE for this inference is summarized in Figure 1.   

The most significant difference in inference from a traditional model is that it is further compiled after the training phase. This compilation transforms the model into an FHE circuit.  After training, the first phase of the compilation is to quantize the model. FHE algorithms struggle to work with float data and require the weights to have an integer representation. The second phase consists of conducting graph lowering. In this phase, the computational graph is revised to replace non-linear activation functions with a polynomial approximation that are compatible with FHE. The third phase fixes bid-width for each of the intermediate inputs. A computational graph is built to compute the bid-width for each step with a range of potential values at run time. This is done to avoid potential overflows. The quantized graph is then passed to a lower level library that bootstraps the circuit operation and ensures it is compatible with FHE. The entire circuit is a model forward pass. The circuit is serialized and produces evaluation keys, a circuit description, and client information \cite{ConcreteMLKeyConcepts2025}. 

At run time during inference, the client encrypts the incoming data and quantized based on the previous compilation. The server loads the complied circuit with the complied arithmetic circuit which contains the evaluation keys that allow the ciphertexts to update. However, the server is not privy to the secret key, thus making it unable to decrypt the information. Homomorphic evaluations are then conducted through each layer with bootstrapping, which avoids issues with noise aggregating. Shuffling is applied throughout the circuit to ensure the security of the ciphertext.Additionally, the output of each intermediate stage remains encrypted. Once the final layer is computed, the class score is returned while still encrypted and is then decrypted by the client and rescaled back into a float \cite{ConcreteMLKeyConcepts2025}.This is then accessed for user inference, allowing the user to interact with it in the application.

\begin{figure}[!ht]
\centering
\begin{tikzpicture}[
  node distance=7mm,
  box/.style={rectangle, rounded corners, draw, align=center, font=\small, minimum width=6cm, fill=gray!5},
  arrow/.style={-{Latex}, thick}
]

\node (train)   [box,] {1) Train plaintext model \\ (normal ML training on raw data)};
\node (compile) [box, below=of train,] {2) Compile with Concrete-ML \\ Quantize weights / activations, generate FHE circuit};
\node (keygen)  [box, below=of compile] {3) Client key generation \\ Secret key kept private, public / evaluation keys shared};
\node (encrypt) [box, below=of keygen] {4) Client encrypts input features \\ using public key};
\node (evaluate)[box, below=of encrypt] {5) Server evaluates compiled circuit \\ ciphertext-in $\rightarrow$ ciphertext-out};
\node (return)  [box, below=of evaluate] {6) Server returns encrypted output};
\node (decrypt) [box, below=of return] {7) Client decrypts result \\ obtains plaintext prediction};

\draw[arrow] (train)   -- (compile);
\draw[arrow] (compile) -- (keygen);
\draw[arrow] (keygen)  -- (encrypt);
\draw[arrow] (encrypt) -- (evaluate);
\draw[arrow] (evaluate)-- (return);
\draw[arrow] (return)  -- (decrypt);

\end{tikzpicture}
\caption{General workflow of FHE inference with Concrete-ML.}
\end{figure}
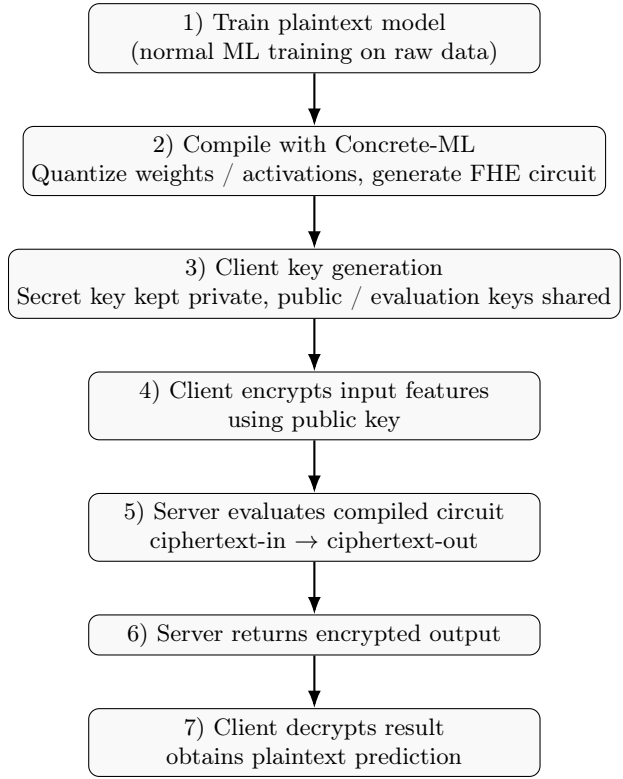

The application is then deployed via a container and further organized through the use of an orchestrator. In this instance, a lightweight flavor of Kubernetes designed for local development (e.g., Minikube) is utilized\cite{minikubeDocs}. A reinforcement learning (RL) agent is then developed to optimize the configuration of the orchestrator based on a simulation, tuning the deployment of replicas to meet operational demands.

\enlargethispage{2\baselineskip}
\section{Experimental Set Up for RL}

As discussed above, we use RL for configuration optimization.
Reinforcement Learning is a paradigm of machine
learning that treats problems as a game. As part of this
formulation, an agent receives rewards and has access to a set
of actions. By learning a strategy, that is a policy, it maximizes
cumulative reward. The agent identifies an optimal way to play
the game to solve the problem.

This is more formally defined through a Markov Decision Process (MDP). In an MDP, the agent interacts with an environment and transitions between different states according to certain probabilities. The policy is the agent's strategy for navigating the environment, which seeks to maximize cumulative reward. The agent learns to anticipate future states based on the dynamics of the environment.

\enlargethispage{2\baselineskip}
\subsection{Environment}
The environment in this specification follows the previously described deployment architecture, comprising a database to cache recommendations, a trained model to generate recommendations, and a Flask app to bridge the services.

This environment is designed around Kubernetes. It is an open-source platform for managing containerized workloads as an orchestration tool \cite{kubernetes_overview}. In Kubernetes, pods are the smallest units of control \cite{kubernetes_pods}. Replicas refer to the desired number of pod instances defined. Essentially, pods are the actual running instances, but replicas represent the target that the system attempts to reach \cite{uma_k8s_resources}.The architecture of the environment is summarized in Figure 2. 

\begin{figure}[htbp!]
    \centering
    \includegraphics[width=1\linewidth]{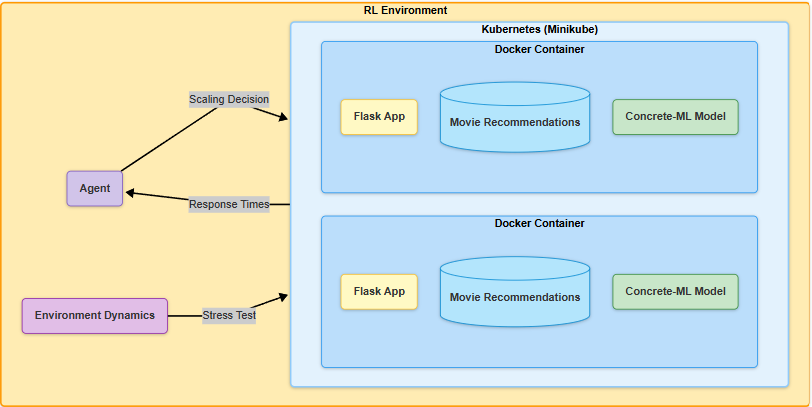}
    \caption{Learning Environment}
    \label{fig:enter-label}
\end{figure}

This environment includes additional dynamics to add realism. Specifically, it provides a stress test for the application by simulating a spike in user demand. This is accomplished by sending a random number of requests, between 5 and 10, concurrently. This load test is intended to overwhelm the system when only a single replica is present. Furthermore, a penalty is applied if an HTTP request does not return a 200 status code or if the response time exceeds two seconds. Additionally, there is a check on the number of replicas: if the agent allocates an unrealistic number of replicas, they are reduced via an additional penalty. In this setting, the maximum number of replicas is 100.

\subsection{Actions Space}
The agent has three actions it can take in this environment: scaling down the number of replicas, maintaining the current number of replicas, or scaling up the number of replicas. The default load balancing algorithm is round robin as utilized by Kubernetes \cite{kubeproxy_doc}. In this strategy, requests are distributed in a turn-taking fashion, with each request being sent in a cycle to balance traffic evenly.

\subsection{State Space}
The state space is defined as a combination of response times and the number of replicas. Since response time is continuous, so is the state space. For learning, the observation space is limited to 1–10 replicas, though the system can scale beyond this for more realistic environment dynamics.

\begin{equation}
 S = \{ (t, p) \mid t \in [0.0, 10.0] \text{ and } p \in \{1, 2, \dots, 10\} \}   
\end{equation}

\subsection{Reward}
The reward structure has two main mechanisms. First, there is a penalty based on the response time for processing requests. If there are insufficient replicas, the agent is encouraged to scale up.

The first part of the reward consists of taking the negative response time to incentivize quicker responses. This value is then multiplied with the number of pods using a scaling factor, which discourages deploying excessive resources. The pod count is used to more accurately reflect the ground truth of the system; there could be a disparity in the replica count due to delays in scaling operations.

\begin{equation}
\text{reward}_{\text{base}} = -\text{response\_time} - 0.1 \times \text{pod\_count}
\end{equation}

To ensure that an excessive number of replicas is not allocated, a secondary reward term is introduced. The maximum allowed number of replicas for the experiment is 100. If the agent attempts to allocate more than 100, a penalty is applied that increases based on the extent to which the allocation exceeds the threshold. Otherwise, no penalty is applied. In effect, the agent is simply trying to optimize the number of replicas in the environment.

The number of replicas is preferred over the number of pods, since they are ephemeral. If a pod dies, Kubernetes will spin up a replacement to maintain the replica count. This is why the agent is manipulating the replicas. 

While the pod count represents the actual number of running pods, the replica count is what the agent controls. Using the replica count for the resource penalty allows for penalizing the agent for over-provisioning resources according to its decisions, in spite of potential lags. Thus, it provides a more reliable signal for dynamic decision-making.

\resizebox{\columnwidth}{!}{$
\text{penalty}_{\text{resource}} =
\begin{cases}
0, & \text{if } \text{current\_replicas} \leq \text{max\_pods}, \\[1ex]
2 \times \left( \text{current\_replicas} - \text{max\_pods} \right), & \text{if } \text{current\_replicas} > \text{max\_pods}.
\end{cases}
$}
\\

A penalty is imposed for failing stress tests. The goal of the stress test is to make the agent robust to unexpected spikes. The success rate is the number of requests that returned 200. If some of the requests failed, the stress test penalty is increased. If the latency is longer than 2 seconds, an additional penalty is imposed. But this penalty is lesser than if the requests failed.
\begin{equation}
\text{penalty}_{\text{stress}} = 5 \, (1 - \text{success\_rate}) + 2 \, \max\{0,\, \text{avg\_response} - 2\}
\end{equation}

The reward can be summarized as such:
\begin{equation}
\text{reward} = \text{reward}_{\text{base}} - \text{penalty}_{\text{stress}} - \text{penalty}_{\text{resource}}
\end{equation}
By taking the response, number of pods, and penalties for robustness and resource conservation, the agent is able to receive a signal to inform scaling.

\subsection{Environment Robustness} To improve the usability of the environment, additional features were integrated throughout the training process. Initially, the agent incorrectly interpreted networking issues as a need for additional resources. For example, if a service went unresponsive, the agent would mistakenly conclude that more replicas were required to meet demand. To address this, self-healing mechanisms and cool-down periods were implemented. Specifically, if a non-200 status code was received and no timeout had occurred in the past 30 seconds, a self-healing routine is triggered. This routine restarts the deployment of the services and cleans up terminated pods as part of garbage collection. If, after self-healing the response code remains invalid, a penalty of -10 is applied.

It is important to note that these self-healing and cool-down features are not part of the action space but are additional dynamics of the environment designed to facilitate robust training. The dynamics of the environment and self-health strategies are summarized below in Algorithm \ref{alg:deployment-health}. 

\newpage

\begin{algorithm}[H]
\caption{Deployment Environment Health Monitoring and Scaling}
\label{alg:deployment-health}

\KwIn{Service URL $U$, initial replicas $r=1$}
\KwIn{Stress interval $k$, resource threshold $r_{\max}=5$}

\While{training}{
  \textbf{Action:} agent selects $a \in \{0: \text{scale down}, 1: \text{no-op}, 2: \text{scale up}\}$\;
  Update replica count $r$ accordingly\;
  Apply scaling to deployment\;
  Wait for system stabilization\;

  \BlankLine
  \tcp{Measure base health}
  $t \gets$ \text{GetResponseTime}{$U$}\;
  $p \gets$ \text{GetPodCount}{}\;

  \BlankLine
  \tcp{Base reward}
  $R \gets -t - 0.1 \cdot p$\;

  \BlankLine
  \tcp{Periodic stress test}
  \If{step counter $\bmod\;k = 0$}{
    penalty $\gets$ \text{StressTest}{$U$}\;
    $R \gets R - penalty$\;
  }

  \BlankLine
  \tcp{Resource usage penalty}
  \If{$p > r_{\max}$}{
    penalty $\gets 2 \cdot (p - r_{\max})$\;
    $R \gets R - penalty$\;
  }

  \BlankLine
  \tcp{Automatic recovery scaling}
  \If{stress penalty $> 0$}{
    $r \gets \min(r + \max(1, penalty / 2), 10)$\;
    Apply scaling to $r$\;
    Wait for stabilization\;
  }

  \BlankLine
  \Return state $[t, p]$, reward $R$\;
}
\end{algorithm}

\FloatBarrier
\subsection{Training}
Proximal Policy Optimization (PPO) is used to train the agent using a multi-layer perceptron to estimate the policy. The "\textbf{stable-baselines}" library is used to train the agent \cite{raffin2021stablebaselines3}. PPO implements a clipping mechanism while doing the updates. This leads to stable training and generally good outcomes. The MLP has a policy and value head. The policy head outputs logits for each of the actions and applies a softmax to transform them into probabilities to create a distribution to determine which actions the agent should take \cite{stablebaselines3_custom_policy}. This is the policy, that is the strategy, the agent will follow. PPO also leverages a ``critic." The critic is a value function that estimates the expected cumulative future reward from a state and determines the advantage: a measure of how much better an action is given in a state compared to the average action in that particular state. It is a scalar value to better inform the policy.

\FloatBarrier

\section{Results}
The agent was trained for 100 episodes in the environment with the reward, latency, and number of replicas captured per episode. The goal of this experiment is to have the agent choose the optimal number of replicas in a production setting. 

In Figure 3, the mean latency and reward of the episodes are grouped by the mean number of replicas. When observing the
reward, it appears that between 3-6 replicas is desirable, since
the reward remains between -1 and -2. In terms of latency,
having two replicas yields the best result in this setting. This is likely in part due to the application taking too long and not being able to handle requests on account of the FHE based inference. The reward is highest around 4 replicas(-1), since it encourages the agent to be defensive. If a usage spike could occur, then the deployment is well prepared. Selecting more replicas is characterized as over-provisioning resources. Through this selecting process, the agent is able to determine a favorable configuration to accommodate the relative sluggishness of FHE based inference while maintaining the performance of the application in the environment.  

\begin{figure}[H]
    \centering
    \includegraphics[width=.94\linewidth]{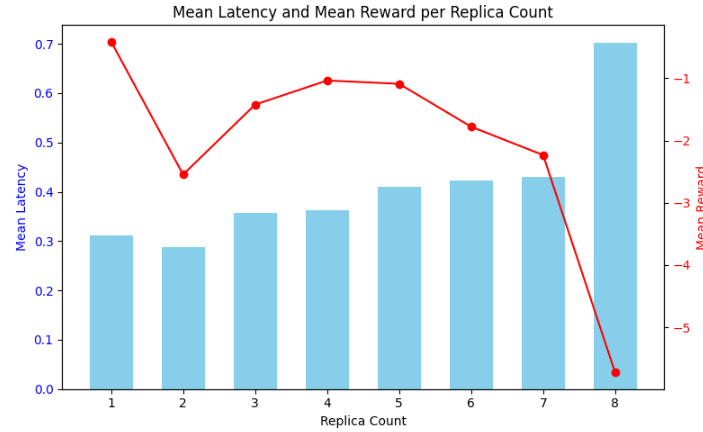}
    \caption{Replica Selection}
    \label{fig:enter-label}
\end{figure}

\newpage

\subsection{Reward}
In Figure 4, the average rewards per episode are observed. 
The agent is attempting to balance having enough replicas to meet expected usage while also handling spikes in user demand. The larger dips in reward (e.g.: 30th step) are likely due, in part, to over-allocation of replicas. The smaller dips (e.g.: 5th step) occur when the agent fails to meet sudden
spikes in demand.
\begin{figure}[H]
    \centering
    \includegraphics[width=1\linewidth]{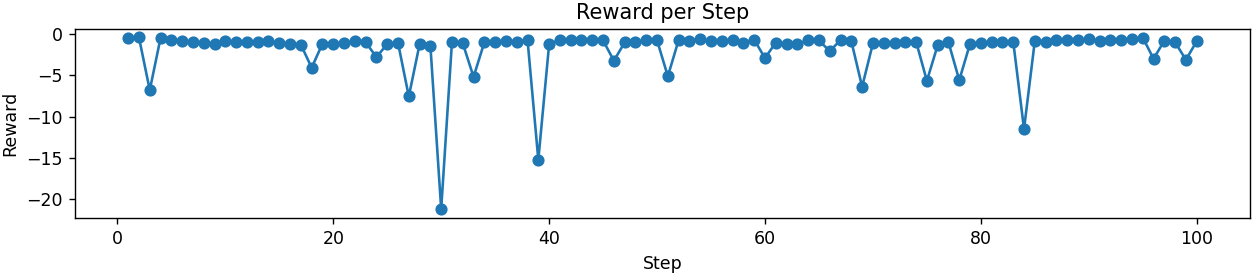}
    \caption{Reward Per Episode}
    \label{fig:reward}
\end{figure}

\subsection{Latency}
In Figure 5, the average latency in seconds per episode is observed. The spikes in latency correspond to the drops in reward, suggesting that the latency component of the penalty is highly informative for the agent.An example of this phenomenon is seen at the 40th step. The average latency appears to be relatively stable and also corresponds to periods when pod are over-provisioned. It may be that the agent overestimates the number of replicas, and the networking delays negate the potential benefits.
\begin{figure}[H]
    \centering
    \includegraphics[width=1\linewidth]{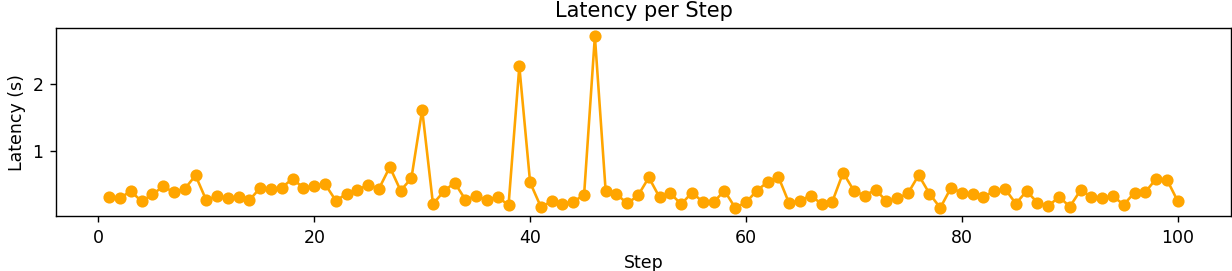}
    \caption{Latency Per Episode}
    \label{fig:latency}
\end{figure}

\newpage

\subsection{Cluster Management}
In Figure 6, the average number of replicas per episode is reviewed.In the early episodes, the number of replicas gradually increases. The increase might be associated with a reward drop at around the 3rd step and higher latency. Eventually, the agent settled on maintaining 3-6 replicas, which appears to be optimal for the given environment.
\begin{figure}[H]
    \centering
    \includegraphics[width=1.04\linewidth]{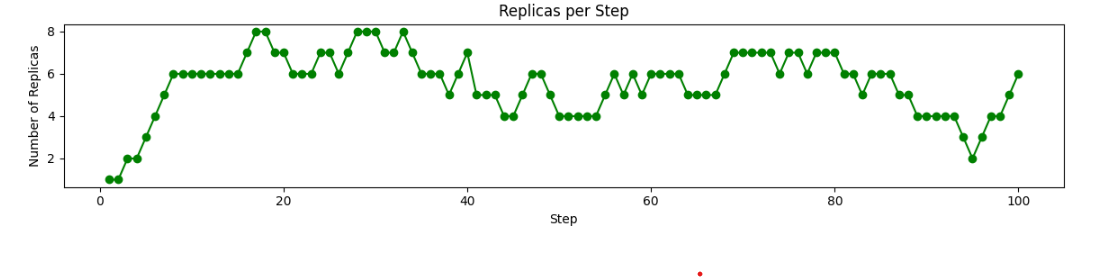}
    \caption{Replicas Per Episode}
    \label{fig:replicas}
\end{figure}

\section{Conclusion}
This work demonstrates that agents can be leveraged to optimize infrastructure and facilitate the deployment of homomorphic applications. By developing a complex environment that integrates Kubernetes with an FHE ecosystem, stronger security guarantees can be achieved while simultaneously mitigating negative impacts on usability.

\section{Future Work}
Future extensions of this work will focus on two main areas. First, the dynamics of the environment can be improved. For example, instead of having a sudden increase in user load, a more gradual ramp-up could allow the agent more time to respond effectively. This is useful in attempting to support deployment of computationally intensive application. By tailoring strategies to fit the deployment dynamics, agents can more effectively derive policies to meet operational demands. Secondly, additional optimizations can be combined. In \cite{marinelli2024}, database indexes were optimized based on application logic for a homomorphic task. A similar approach could be applied to the databases, treating infrastructure optimizations as a curriculum learning task. The environment is located at \hyperlink{https://anonymous.4open.science/r/movie\_app-B3F6/README.md}{https://anonymous.4open.science/r/movie\_app-B3F6/README.md}

\newpage
\bibliographystyle{unsrt}
\bibliography{references}    

\end{document}